\documentclass[twocolumn, twocolappendix]{aastex631}

\usepackage{multirow}
\usepackage{natbib}
\usepackage{amsmath}

\graphicspath{{./}{fig/}}

\newcommand{\Hb}{H$\beta$}

\newcommand{\MgII}{Mg\,\textsc{II}}
\newcommand{\OIII}{[O\,\textsc{III}]}

\newcommand{\CIV}{C\,\textsc{IV}}

\newcommand{\FeII}{Fe\,\textsc{II}}

\newcommand{\ergs}{\text{erg s$^{-1}$}}

\newcommand{\flux}{\text{erg cm$^{-2}$ s$^{-1}$}}

\newcommand{\kms}{\text{km s$^{-1}$}}

\newcommand{\gaa}{\text{\AA}}

\newcommand{\SDSSIV}{SDSS \uppercase\expandafter{\romannumeral4}}

\defcitealias{Ren2022}{Paper I}

\hypersetup{
   colorlinks,
   linkcolor={blue!88!black!80},
   citecolor={blue!88!black!80},
   urlcolor={blue!88!black!80}}
   
\received{September 10, 2023}
\revised{November 26, 2023}
\accepted{December 14, 2023}
\submitjournal{ApJ}

\shorttitle{Multi-epoch spectra of EVQs}
\shortauthors{Ren W. et al.}

\graphicspath{{./}{figures/}}

\begin{document}

\title{Extreme Variability Quasars in Their Various States. II: Spectral Variation Revealed with Multi-epoch Spectra}

\correspondingauthor{Wenke Ren, Junxian Wang}
\email{rwk@mail.ustc.edu.cn, jxw@ustc.edu.cn}

\author[0000-0002-3742-6609]{Wenke Ren}
\affiliation{CAS Key Laboratory for Research in Galaxies and Cosmology, Department of Astronomy, University of Science and Technology of China, Hefei, Anhui 230026, China}
\affiliation{School of Astronomy and Space Science, University of Science and Technology of China, Hefei 230026, China}

\author[0000-0002-4419-6434]{Junxian Wang}
\affiliation{CAS Key Laboratory for Research in Galaxies and Cosmology, Department of Astronomy, University of Science and Technology of China, Hefei, Anhui 230026, China}
\affiliation{School of Astronomy and Space Science, University of Science and Technology of China, Hefei 230026, China}

\author[0000-0002-4223-2198]{Zhenyi Cai}
\affiliation{CAS Key Laboratory for Research in Galaxies and Cosmology, Department of Astronomy, University of Science and Technology of China, Hefei, Anhui 230026, China}
\affiliation{School of Astronomy and Space Science, University of Science and Technology of China, Hefei 230026, China}

\author{Xufan Hu}
\affiliation{CAS Key Laboratory for Research in Galaxies and Cosmology, Department of Astronomy, University of Science and Technology of China, Hefei, Anhui 230026, China}
\affiliation{School of Astronomy and Space Science, University of Science and Technology of China, Hefei 230026, China}

\begin{abstract}
We previously built a sample of 14,012 extremely variable quasars (EVQs) based on SDSS and Pan-STARRS1 photometric observations. In this work we present the spectral fitting to their SDSS spectra, and study the spectral variation in 1,259 EVQs with multi-epoch SDSS spectra (after prudently excluding spectra with potentially unreliable spectroscopic photometry). We find clear ``bluer-when-brighter" trend in EVQs, consistent with previous findings of normal quasars and AGNs. We detect significant intrinsic Baldwin effect (iBeff, i.e., smaller line EW at higher continuum flux in individual AGNs) in the broad \MgII\ and \CIV\ lines of EVQs. Meanwhile, no systematical iBeff is found for the broad \Hb\ line, which could be attributed to strong host contamination at longer wavelengths. Remarkably, by comparing the iBeff slope of EVQs with archived changing-look quasars (CLQs), we show that the CLQs identified in literature are mostly likely a biased (due to its definition) sub-population of EVQs, rather than a distinct population of quasars. We also found no significant broad line breathing of either \Hb, \MgII\ or \CIV, suggesting the broad line breathing in quasars may disappear at longer timescales ($\sim$ 3000 days). 
\end{abstract}



\section{Introduction} \label{sec:intro}

One of the hallmarks of quasars is their universal variability across all wavelengths. Quasar light curves typically exhibit variations with an amplitude of $\sim$0.2 mag in the rest-frame UV/optical band, occurring on timescales of months to years \citep[e.g., ][]{VandenBerk2004, Sesar2007, MacLeod2010}. These variations are often described by a damped random walk (DRW) model first proposed by \citet{Kelly2009}. Reverberation mapping (RM) studies have revealed strong correlations between the variation of the continuum and the broad emission lines (BELs) \citep[e.g.,][]{Grier2017, Grier2019, Homayouni2020}, suggesting that the optical-UV variability of quasars is driven by the innermost region of the accretion disk \citep[e.g.,][]{Ross2018}. However, due to the stochastic and complex nature of quasar variability \citep[e.g.][]{Cai2018a,Sun2020}, the physical processes underlying these variations are still a topic of debate.

In the cutting-edge field of quasar variability research, the phenomenon of changing-look (CL) is particularly intriguing. It has long been noted that a few AGNs exhibit drastic changes in their broad emission lines, with their spectral types transitioning between Type 1 and Type 2 \citep[e.g., ][]{Cohen1986, Goodrich1995}. Initially, due to limited data sizes and time spans, the discovered CLQs are very rare and usually transient. The observed changes in broad lines were often attributed to dust extinction \citep{Risaliti2007}, tidal disruption events, or supernovae \citep{Aretxaga1999a}. However, with the accumulation of data, an increasing number of CLQs ($\sim$100) have been identified \citep[e.g., ][]{Ruan2016, MacLeod2019, Green2022}. Most CL phenomenon are accompanied by extreme continuum variations, and their new spectral states can keep for months to years.  These observational results indicate that the CL phenomenon should be driven by intrinsic but violent, accretion dominated activity. 
What are the underlying mechanisms for such violent process? Are they due to possible accretion transition in analogy to X-ray binaries \citep[e.g.][]{Ruan2019,Yang2023}, or just more violent version of normal variability?
Much larger samples of sources with violent variability are desired to distinguish such possibilities.

Since the identification of CLQs relies on repeated spectral observations showing emerging/disappearing of broad emission lines, which are relatively rare, some related studies initially focus on quasars with extremely strong continuum variability \citep{MacLeod2016, Rumbaugh2018, Sheng2020}. 
Compared with CLQs, much larger samples of such quasars could be easily built solely based on broadband photometric variability, and such samples could provide essential clues to the physical nature of violent variability and its relation with changing-look and normal (less extreme) variability. 

Through systematic research on extremely variable quasars (EVQs), \citet{Rumbaugh2018} found that the equivalent width (EW) of the broad emission lines of EVQs are larger than that of normal quasars. Note \citet{Kang2021} further discovered a general correlation between the variability and emission line EW in large sample of quasars. It is clear that quasar variability has an underlying influence on the broad-line regions (BLRs) and thus is worth investing more efforts.
Furthermore, studies with small samples of AGNs have illustrated that emission lines change along with continuum variations, i.e., the EW of broad lines decreases as the continuum increases \citep[intrinsic Baldwin effect, iBeff, e.g.][]{Goad2004,Rakic2017}; the full width at half maximum (FWHM) of different lines also exhibits varying correlations with continuum luminosity \citep[e.g.][]{Wang2020}. However, restricted by the signal-to-noise ratio (SNR) of available survey spectra, such changes are often not eminent enough to distinguish from noise when extending to quasars, as most of which exhibit low intrinsic variability. In this sense, studying the emission line variability in large samples of EVQs could also shed new light on the BLR physics.
 
In our previous work \citet{Ren2022} \citepalias[hereafter, ][]{Ren2022}, we constructed a sample of EVQs consisting of 14,012 sources using photometric data from the Sloan Digital Sky Survey (SDSS) and Pan-STARRS (PS1). We obtained the composite spectra of the EVQs in various states (according to the relative brightness of each EVQ during SDSS spectroscopic observation compared with the mean brightness from multi epoch photometric observations) and of control samples with matched redshift, luminosity and black hole mass. We verified that EVQs tend to have stronger emission lines in \Hb, \OIII, \MgII, and \CIV. The EW changes between different states shows a clear iBeff in \MgII\ and \CIV, but not in \Hb. Furthermore, we found that EVQs have stronger broad line wings compared to normal quasars, suggesting that EVQs could launch more gas from the inner disk into the very broad line region (VBLR).

In this work, we focus on the EVQs with multi-epoch SDSS spectra to explore the spectral variations in individual EVQs. Based on the EVQ sample we constructed in \citetalias{Ren2022}, we select a sub-sample with repeated spectral observations in SDSS. The structure of this paper is as follows. In \S\ref{sec:data}, we describe the sample selection and methods of spectral measurement. The results are presented in \S\ref{sec:result}. We then discuss our results in \S\ref{sec:discussion} and provide a summary in \S\ref{sec:conclusion}. Throughout this paper, we adopt a flat $\rm{\Lambda}$CDM cosmology with $\Omega_\Lambda=0.7$, $\Omega_m=0.3$, and $H_0=70~\rm{km~s^{-1}~Mpc^{-1}}$.


\section{The Data and Sample} \label{sec:data}

\subsection{The EVQ Sample} \label{subsec:sample}

We start from the EVQ catalog constructed by \citetalias{Ren2022}, which is selected from the SDSS data release 14 (DR14) quasar catalog \citep[DR14Q, ][]{Paris2018}.
Here we provide a brief overview of the selection criteria and spectra fitting procedure. 
Readers are advised to refer to \citetalias{Ren2022} for detailed description. 
Based on all photometric observations ($g$ and $r$) from SDSS and PS1 database for each quasar in DR14Q, we amend the minor differences in filter transmission between two surveys,  reject epochs with potentially unreliable photometric measurements, and build clean light curves for every source. Sources with robust detection of extreme variability in both $g$ and $r$ light curves  (with $\Delta g_{max} > 1$ and $\Delta r_{max} > 0.8$) are selected as EVQs. In total, we select 14,012 EVQs with 20,069 archived SDSS spectra. Each spectrum is processed using {\tt PyQSOFit} \citep{2018ascl.soft09008G} following the procedure described in \citetalias{Ren2022}. The catalog of the EVQs and the spectral fitting results are now presented in Table \ref{tab:catalog} of Appendix \ref{B_catalog}.

Among the 14,012 EVQs, 1354 sources have repeated spectral observations in SDSS archive with a continuum SNR $>3$. In Appendix \S\ref{A_drw}, we develop a method to identify spectra with potentially unreliable spectroscopic photometry (such as due to fiber-drop) by comparing the observed spectroscopic photometry with the expected value derived through  modeling the photometric light curves with a damped random walk (DRW) process. After excluding such possibly dubious spectra, 1259 multi-spectra EVQs are remained for further analyses.

\subsection{Spectral measurements} \label{subsec:specmeasure}

The spectra are fitted following \citet{Shen2011} using {\tt PyQSOFit} \citep{2018ascl.soft09008G}. For each spectrum, we fit with a power-law ($f_\lambda \propto \lambda^{\alpha}$) and a broadened \FeII\ template for continuum within tens of separated emission line-free windows and give the measurement of continuum and \FeII\ properties. Then, we cut the spectrum around three luminous broad emission lines: \Hb, \MgII, and \CIV\ if they are within the spectral coverage and then fit the lines locally. The continuum is re-fitted using the windows on both sides of the emission lines. After detracting the continuum, we use a few groups of Gaussian to fit the residuals: 4 Gaussians (2 cores + 2 wings) for \OIII\ doublet, 1 for narrow \Hb\ and three for broad \Hb\ in \Hb\ region; 1 Gaussian for narrow and 3 for broad \MgII\ region; and 3 Gaussians for only broad \CIV\ component in \CIV\ region. 
The detailed procedure and parameter settings can be found in \citetalias{Ren2022}.

For each emission line component, we calculate the fitted peak wavelength ({\tt PEAK}), integrated flux ({\tt FLUX}), equivalent width ({\tt EW}), line dispersion (second moment, {\tt SIGMA}), and full width at half maximum ({\tt FWHM}). Since the broad component of \Hb, \MgII, and \CIV\  (consisting of triple Gaussians) may have asymmetric shapes indicating complex structures of BLRs, we further measure the following properties. The bisectional center \citep[{\tt BISECT,}][]{Sun2018a}, where the wavelength splits the line into two equal areas in flux, has been used in \citetalias{Ren2022} to qualitatively reflect the wing structures in a broad line. To further parameterize the shape and asymmetry of the broad component, we adopt the: i) full width at half maximum ({\tt FWHM}), ii) full width at quarter maximum ({\tt FWQM}), and iii) full width at 10\% maximum ({\tt FW10M}) to denote the BLRs from different distance to the SMBH. We also measure the corresponding shifts of the centers of the above three widths, {\tt Z50}, {\tt Z25}, and {\tt Z10} to show the intrinsic gravitational redshift or inflow/outflow of BLRs (cf. Fig. 6 in \citealt{Rakic2022}).

We employ a Monte Carlo approach to assess the statistical uncertainties of the parameters we measured. This is done through generating 50 mock spectra by adding Gaussian noise to each original spectrum using the reported flux density errors, fitting the mock spectra with the same routines, and deriving the scatter of each parameter.

\begin{figure}[tb!]
   \includegraphics[width=.48\textwidth]{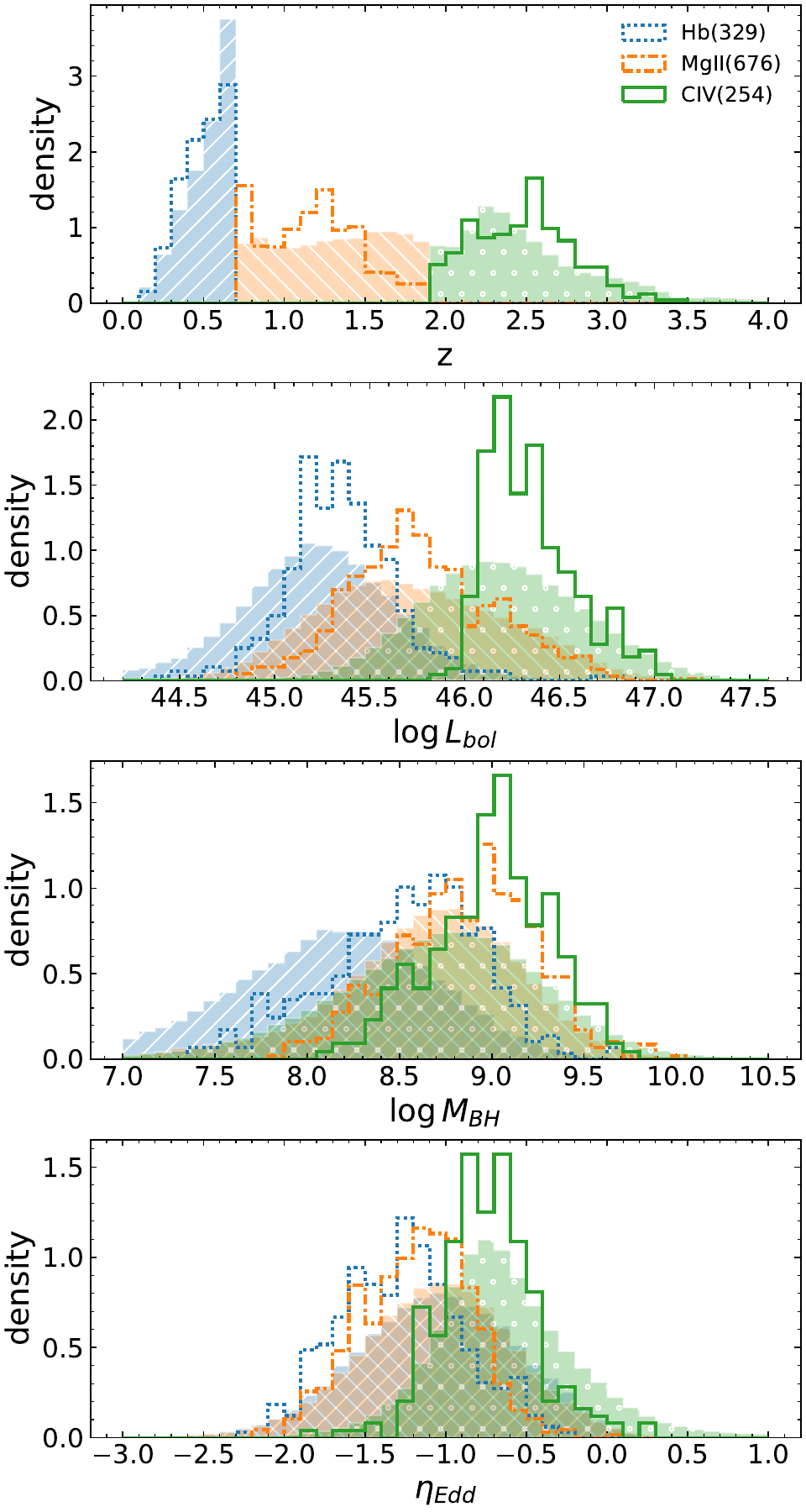}
   \caption{The redshift, $L_{bol}$, $M_{BH}$, and $\eta_{Edd}$ distribution of multi-spectra EVQs. The mean values from multiple spectra are adopted for each EVQ. We split the sample into three groups according to redshift. The source number of each group is noted in the parenthesis in the legend. As a comparison, we also plot the re-normalized distribution of SDSS DR14Q \citep{Rakic2022}, divided by the same redshift criterion and drawn with corresponding colored shades. 
   \label{fig:overallprop}}
\end{figure}

We adopt the bolometric corrections (BCs) and the calibration parameters in \cite{Shen2011} to estimate the bolometric luminosity $L_{bol}$ and the BH mass $M_{BH}$ of a single-epoch spectrum. We take fiducial $L_{bol}$ and $M_{BH}$ calculated from $L_{5100}$ and \Hb\ for sources at $0.08\leq z<0.7$, from $L_{3000}$ and \MgII\ at $0.7\leq z<1.9$, and from $L_{3000}$ and \MgII\ at $1.9\leq z<4.0$ respectively. Our following results will be based on such three sub-samples. The Eddington ratio is also given in our catalog. We adopt the mean value from multiple-epoch measurements for each target and plot the general physical property distribution of multi-spectra EVQs in Fig. \ref{fig:overallprop}.

%


\section{Results} \label{sec:result}

\begin{figure*}[tb!]
   \includegraphics[width=1\textwidth]{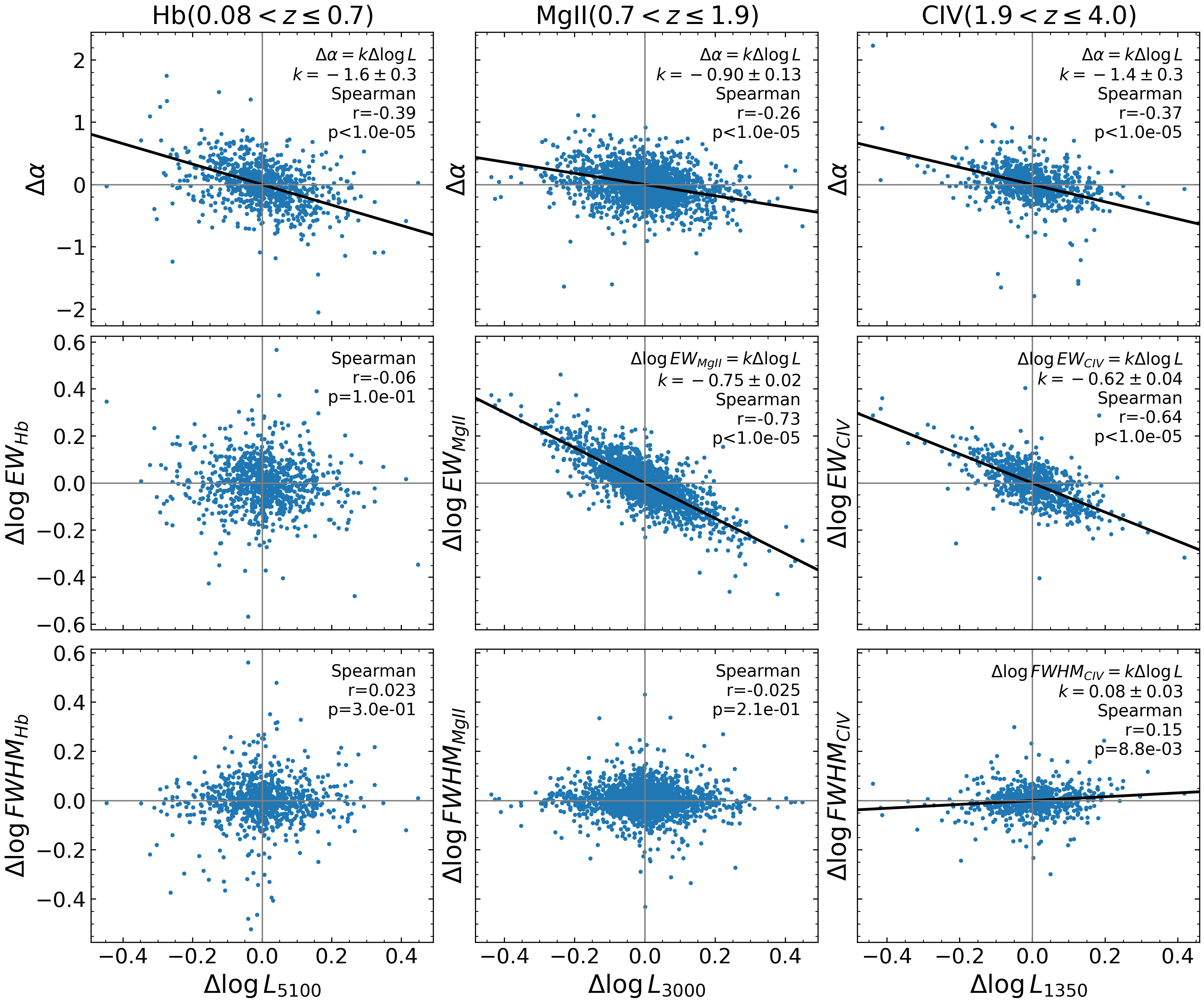}
   \caption{The correlation between the variability of spectral properties and that of continuum luminosity. The variation of continuum power-law slope ($\alpha$), broad line EW and FWHM are shown in rows sequentially. The variability for each parameter is derived through subtracting the mean value of multi-epoch spectral measurements for the same EVQ from a single-epoch measurement. The Spearman correlation coefficients and the p-values (derived through bootstrapping the sample, hereafter the same, see the 2nd parapha of \S\ref{sec:result}  for details) are shown in the upper right of each plot. For correlations with p-value$<$0.01, we use a black solid line to show the orthogonal distance regression result and present the derived regression slope $k$ with 1$\sigma$ uncertainties derived through bootstrapping (hereafter the same).
   \label{fig:Delta_L}}
\end{figure*}

In this section, we explore how the spectral properties of EVQs vary along with the continuum flux, in three distinct samples split by redshift to guarantee the emission line measurements. In total, we have 329 sources with 920 spectra for the \Hb\ sample, 676 sources with 2206 spectra for the \MgII\ sample, and 254 sources with 799 spectra for the \CIV\ sample. About $2/3$ of the objects have only two spectra, and there are 15 sources with more than 20 observations which are all from the SDSS-RM program \citep{Shen2015}.

To present sources with different luminosities together and explore their average variability, our results are primarily expressed in terms of variation. To fully leverage each observation, we derive the deviation of each single-epoch measurement from the mean value for each EVQ. In this case, one single-epoch spectrum will contribute one data point to our following plots. We note that the various data points from the same EVQ are thus not completely independent to each other (since calculating the mean value consumes one degree of freedom for each EVQ) and the traditional p-value of the Spearman correlation evaluated from Student's t-test is no longer valid. Therefore, we estimate the p-value by bootstrapping the quasar sample instead. Note since we run bootstrap 100,000 times, thus we are unable to give exact p-value smaller than 1.e-5.


\subsection{Color variation} \label{subsec:color}

The so-called ``bluer-when-brighter'' (BWB) pattern has been widely seen in quasars and AGNs \citep[e.g.][]{Trevese2002, Sun2014, Guo2016}. In \citetalias{Ren2022}, the composite spectra of EVQs indirectly support such a pattern. We quantitatively test this conclusion with the multi-spectra EVQs by exploring the spectral slope variability in individual sources.

The first row in Fig. \ref{fig:Delta_L} shows the changes of spectral slope versus the changes of log-luminosity. The anti-correlation between $\Delta \alpha$ and $\Delta \log{L}$ is found in all three samples, denoting a consistent BWB pattern up to $z\sim4$. We use the orthogonal distance regression \citep{Boggs1989} method provided by {\tt scipy.odr} package to fit the data with the data errors considered (hereafter the same, unless otherwise stated). The regression slope $k$ for the three samples are $-1.6\pm0.3$, $-0.90\pm0.12$, and $-1.4\pm0.3$ ((with the 1$\sigma$ statistical uncertainties derived from bootstrapping the samples, hereafter the same), respectively, roughly consistent with the results from PG QSOs \citep[e.g.][]{Pu2006}.

The slightly steeper slope from the \Hb\ sample could be due to the contamination of the host galaxy as we do not decompose the host galaxy component while fitting the continuum. As a result, for those low-z sources for which the host contamination could be more prominent, the $\Delta \log{L_{5100}}$ will be underestimated, and the $\alpha$ at dim states will be overestimated. Besides, the relative flux of the host galaxy component may also vary depending on the observation conditions \citep{Zhang2013, Hu2015}. Nevertheless, given their high luminosity as quasars (see Fig. \ref{fig:overallprop}) and the strong correlation at higher redshift, the host alone should not be fully responsible for the observed spectral changes. Indeed, it has been well demonstrated that the host contamination can not dominate the observed color variation in SDSS quasars, as host contamination can not reproduce the observed timescale-dependence of the color variation \citep{Sun2014,Cai2016,Zhu2016}.


\subsection{Variation of line EW and FWHM} \label{subsec:EW}

In the second row of Fig. \ref{fig:Delta_L}, we explore the iBeff of the three most luminous broad lines. Strong anti-correlations between the variation of line EW and that of the continuum luminosity are found in \MgII\ and \CIV\ while only marginal anti-correlation in \Hb. The averaged iBeff slope for \MgII\ and \CIV\ are comparable but slightly steeper than the results revealed with the stacked spectra in \citetalias{Ren2022} and those from  literature \citep{Pogge1992, Kong2006, Homan2020, Zajacek2020}. The steeper iBeff of \MgII\ and \CIV\ could likely be attributed to the large variability of our sample, as \cite{Homan2020} also found a steeper slope in their more variable sample. 

We note that the missing anti-correlation between $EW_{H\beta}$ and $L_{5100}$ does not means the absence of iBeff in \Hb, as the host component has not been properly considered in our spectral fitting. The blend of the host galaxy in the spectra causes significant underestimation of $EW_{H\beta}$, particularly in dim states, thereby weakening the underlying anti-correlations. Subtracting the host component from such low SNR individual spectra, which is beyond the scope of this work, would be highly unreliable and would introduce great uncertainties to spectral measurements \citep[e.g.][]{Wu2022}. 

\begin{figure}[tb!]
   \includegraphics[width=0.46\textwidth]{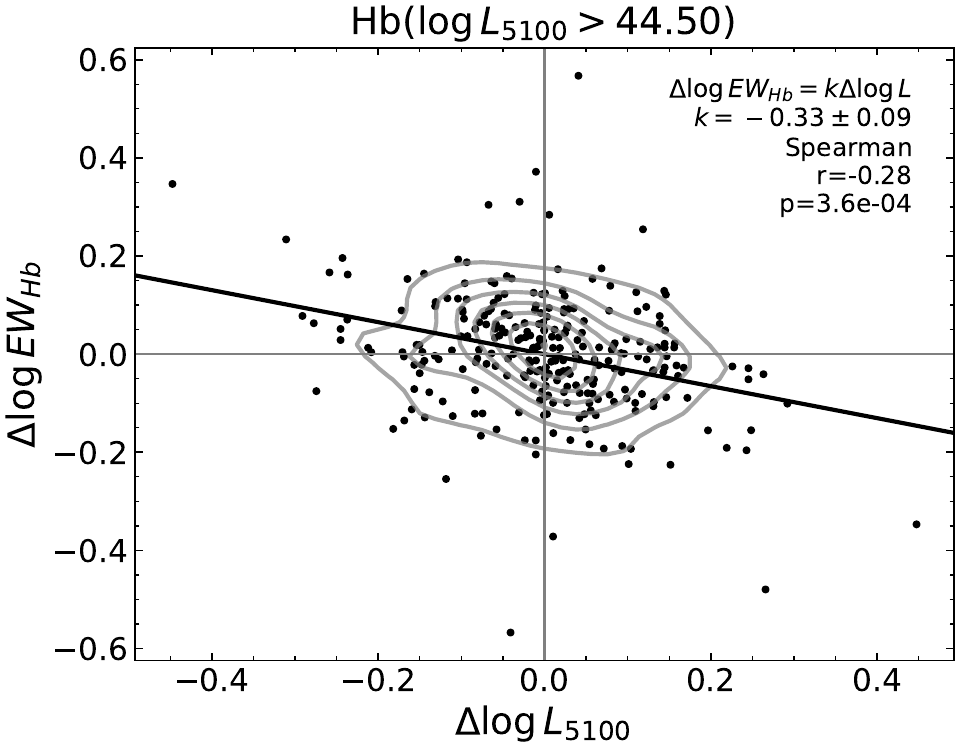}
   \caption{The broad \Hb\ iBeff is visible in a sub-sample of more luminous EVQs, suggesting the absence of broad \Hb\ iBeff shown in the lower left panel of Fig. \ref{fig:Delta_L} is due to host contamination. The labels are similar to those in Fig. \ref{fig:Delta_L}.
   \label{fig:HighL_iBeff}}
\end{figure}

To alleviate the host contamination, we build a luminous sub-sample from our EVQs which contains about 1/3 of the total sample with $\log \overline{L}_{5100}>44.5$. The broad \Hb\ iBeff of this sub-sample is shown in Fig. \ref{fig:HighL_iBeff}. A robust negative correlation is observed in this sub-sample, indicating that the disappearance of the iBeff of \Hb\ line in the full sample is likely due to the host contamination. We would further discuss the effects of host contamination in \S\ref{subsec:CLQs}, where the connection between the EVQs and CLQs are investigated.

We also explore the so-called "breathing" of the broad lines, e.g., how the line width changes with the luminosity (see the third row of Fig. \ref{fig:Delta_L}). Only weak and marginal anti-breathing (positive correlation) is found in the \CIV\ sample, and no clear sign of breathing is observed in the \MgII\ sample. These results are generally consistent to those reported in \citet{Wang2020}. However, we find that the \Hb\ sample also shows no breathing according to Spearman test, contradicting previous findings \citep[e.g., ][]{Denney2009, Park2012, Wang2020}. This paradox will be discussed in \S\ref{subsec:breathing}.


\subsection{Line profile and asymmetry} \label{subsec:shape}

\begin{figure*}[tb!]
   \includegraphics[width=0.95\textwidth]{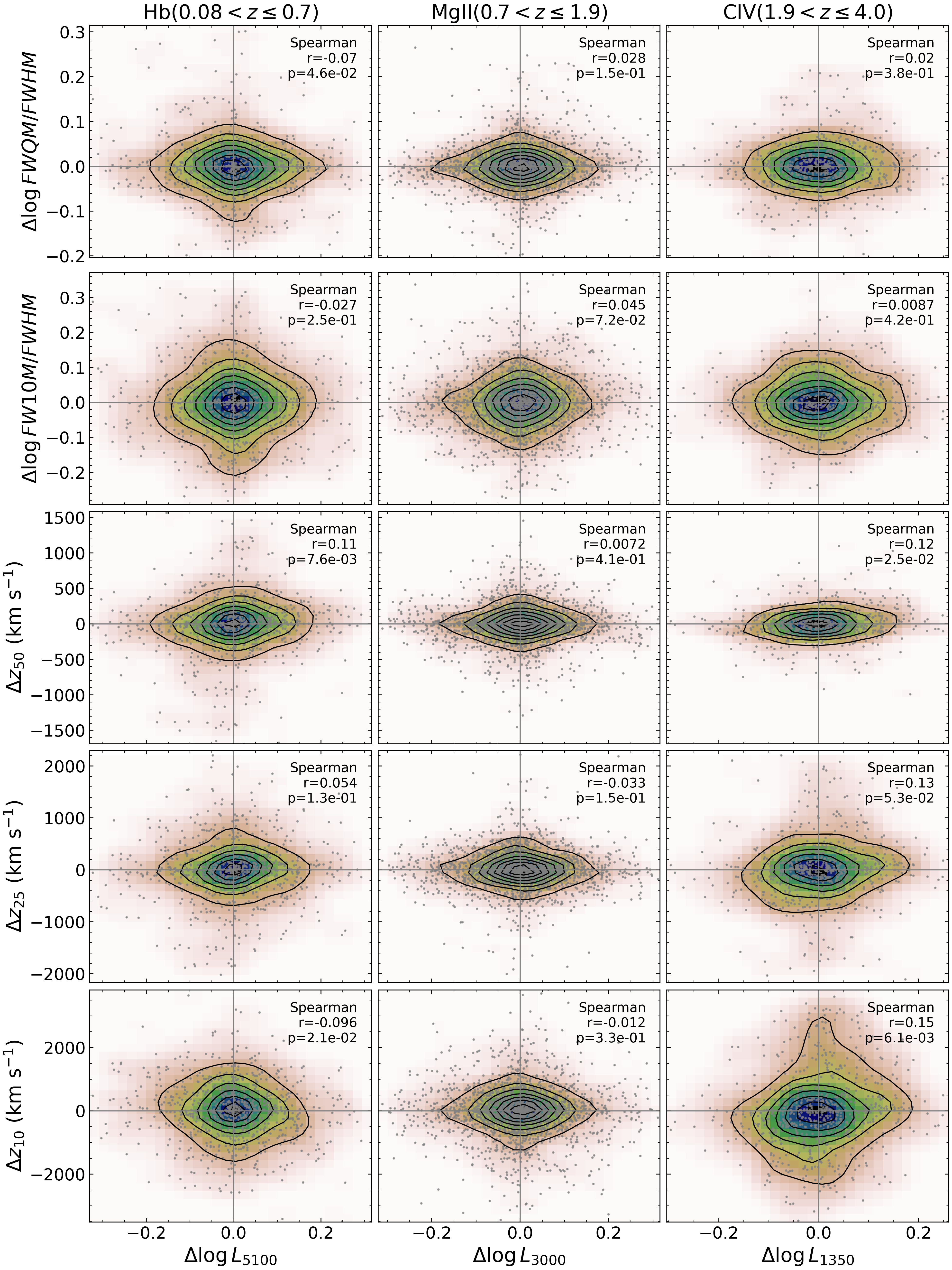}
   \caption{The variation of broad line profile and asymmetry (see text for description) versus that of continuum luminosity. The labels are similar to those in Fig \ref{fig:Delta_L}
   \label{fig:Delta_shape}}
\end{figure*}

In \citetalias{Ren2022}, we found that the EVQs have slightly stronger broad wings compared with control samples. Here we investigate whether the broad line profiles vary with continuum flux in EVQs with multi-epoch spectra. We use the line width ratio ({\tt FWQM/FWHM} and {\tt FW10M/FWHM}) to denote the line profile and the shift of line center ({\tt Z50}, {\tt Z25}, and {\tt Z10}) to quantify the line asymmetry. The changes of above parameters with the continuum luminosity are shown in Fig. \ref{fig:Delta_shape}. 

According to the Spearman correlation coefficient and its p-value, we do not find any strong dependency between line profiles and continuum flux. This suggests the BLR structure does not systematically change with continuum variation. Such a result also indicates that the EVQs are not an unusual population with exceptionally large variability.


\section{Discussion} \label{sec:discussion}

\subsection{EVQs and CLQs} \label{subsec:CLQs}

In \S\ref{subsec:EW} we found the \Hb\ EW of EVQs does not vary with continuum flux (but with considerable scatter), i.e., showing no clear iBeff on average. In such cases, there will naturally be some EVQs that have lower \Hb\ EW in their noisy dim-state spectra which may yield the so-called ``vanishment'' of broad Balmer lines (i.e., CLQs). In this section, through this entry point, we will delve deeper into whether CLQ is a distinct population or simply a biased sub-population of EVQ.

\subsubsection{Archived CLQs}

The study of CLQs has been heating up in recent years, decades after the discovery of Changing-look AGNs \citep[e.g., ][]{Tholine1976, Cohen1986}. So far, nearly a hundred of CLQs have been claimed, but a unique classification is still lacking. The term ``Changing-look'' is an iconic description (highly qualitative but not quantitative) firstly used to denote the appearance or disappearance of broad emission lines. Since the CL phenomenon is relatively rare at the early stage, it was feasible to use AGN types or merely by visual checks to identify CL-AGNs. Such subjective identification method is inherited to the research of CLQs.

Since the CL phenomenon is usually accompanied by dramatic continuum changes, one developed approach to efficiently search for CLQs is to find candidates from quasars with strong continuum variation, and then pick CLQs through visual check \citep[e.g.][]{MacLeod2016}. Obviously, the process of visual check can be easily influenced by individual subjective judgements.

With the accumulation of multi-epoch spectra, researchers have explored another method for hunting CLQs. For the spectra of CLQs in dim states usually have no broad lines, they were often classified as galaxies by the initial pipelines. Therefore, one can search for CLQs by directly examining sources with repeated spectra but with different classifications \citep[e.g.][]{Ruan2016, Yang2018}. Such semi-quantitatively search for CLQs is highly dependent to the classification pipeline, and in many cases still dependent on visual inspection \citep[e.g., 19 confirmed CLQs out of 10,204 candidates by visual check in][]{Yang2018}.

In order to further standardize the selection of CLQs, a few recent works start to focus on the change of emission line flux itself. \citet{MacLeod2019} proposed that a quasar could be deemed as a CLQ if the SNR (signal-to-noise ratio) of the broad line flux change between two epochs is greater than 3. Meanwhile \citet{Yang2018} and \citet{Sheng2020} directly use the SNR of broad line to decide whether a broad component is apparent or not. The most recent work by \citet{Green2022} combined the quantitative criteria and visual inspection. These efforts indeed make the search for CLQs more objective and consistent. However, the criterion above highly rely on the SNR of the spectra, and they are unstable when applying to spectra observed in different conditions and could be sensitive to the quasar luminosity. For example, as \citet{MacLeod2019} have noted that their methods may select quasars with minor variation in \Hb\ line flux when their spectra have very high SNR. Furthermore, applying quantitative criteria on the broad line flux is not straightforward, as the broad line flux is expected to vary with continuum in all quasars and AGNs (the cornerstone of reverberation mapping studies). 

To explore whether there are intrinsic differences between currently archived CLQs and EVQs, we collected all archived \Hb\ CLQs and refitted their spectra using the same method we applied to our EVQs. Note in literature there are only 3 MgII CLQs and 3 \CIV\ CLQs claimed \citep[][see also \S\ref{subsec:biasedpopulation}]{Guo2019,Guo2020a,Ross2020}, thus here we focus on Hb select CLQs only. We gathered a total of 79 securely-identified CLQs from \citet{LaMassa2015a, Ruan2016, MacLeod2016, Yang2018, MacLeod2019, Sheng2020, Green2022, Lopez-Navas2022}. To ensure the consistency of analyses, we only retained CLQs with repeated SDSS spectra in both dim and bright states. As a result, 36 sources are preserved, all with $z<0.65$. Among those CLQs, 14 are included in  our EVQ sample. The other 22 CLQs do not satisfy our strict selection criteria of EVQs (i.e., $\Delta g_{max}$ $>$ 1 mag and $\Delta r_{max}$ $>$ 0.8 mag, in SDSS \& PS1 photometric light curves well cleaned against potential photometric defects, see \citealt{Ren2022}).

Indeed, our independent spectral fitting reveal broad \Hb\ lines in the dim states of 33 CLQs, indicating in most archived CLQs the broad \Hb\ indeed does not completely disappear in the dim state. In the remaining three cases, two of them have their \Hb\ lines in the dim states completely submerged by the host galaxy, leaving an obvious absorption feature. The dim-state spectrum of the third CLQ is flawed without enough wavelength coverage around \Hb. These 3 dim-state spectra are therefore excluded from further analysis below. 

\subsubsection{Comparing the iBeff of EVQs with CLQs} \label{subsubsec:host}

\begin{figure}[tb!]
   \includegraphics[width=0.48\textwidth]{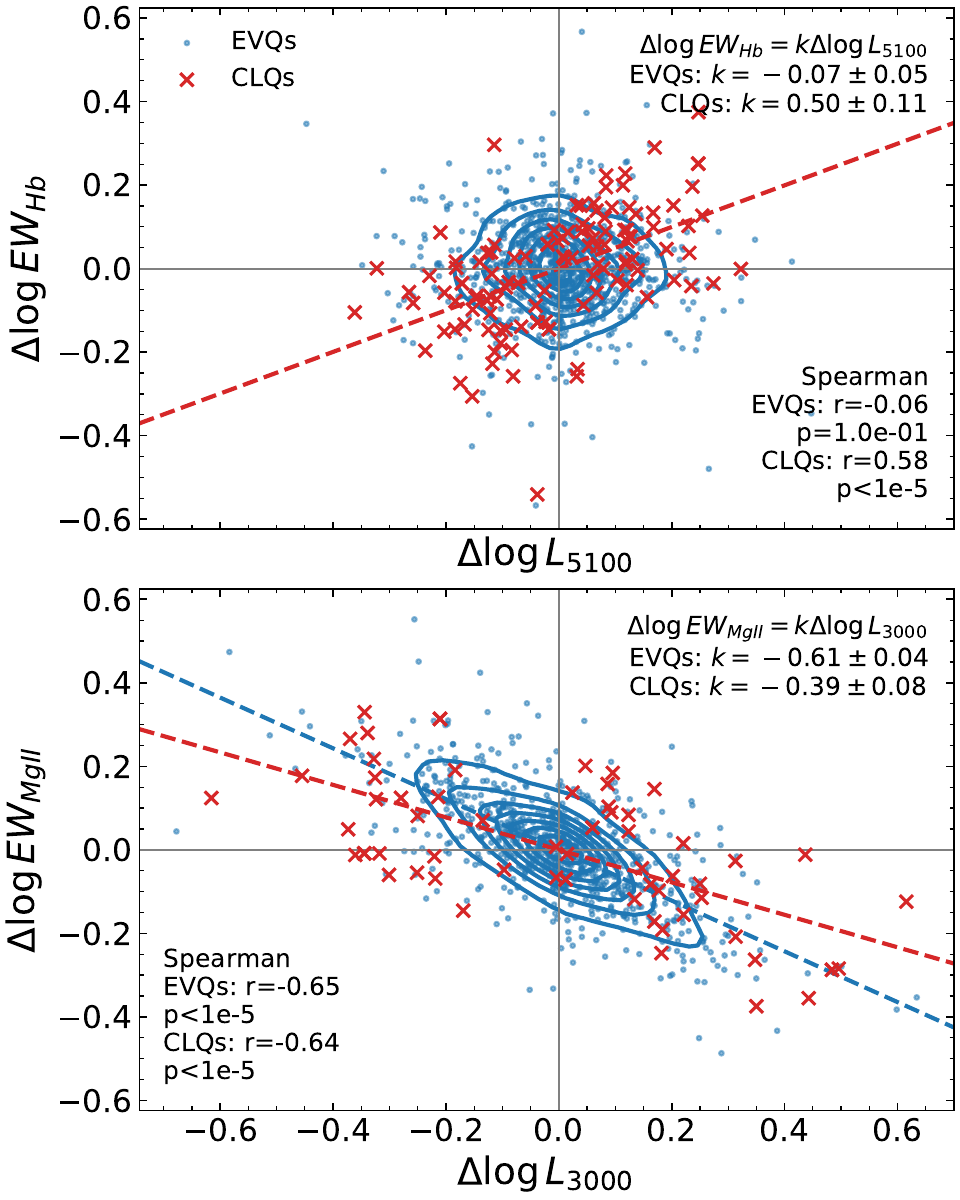}
   \caption{The orthogonal distance regression correlation between the changes of continuum luminosity and the changes of line EW. The variation of continuum luminosity and line EW are calculated by subtracting the mean value of each object from the measurement of each epoch, which is identical to that in Fig. \ref{fig:Delta_L}. Upper: the iBeff of \Hb\ of our EVQs and CLQs collected from literature. Lower: the \MgII\ iBeff for the sources plotted in the upper panel but with \MgII\ line measurements. The fitting results and the Spearman correlation coefficient $r$ (with p-values) of each sample is listed. For correlations with p-value$<$0.01, we use a black solid line to show the orthogonal distance regression fitting result.
   \label{fig:CLQ_EW}}
\end{figure}
\begin{figure}[tb!]
   \includegraphics[width=0.48\textwidth]{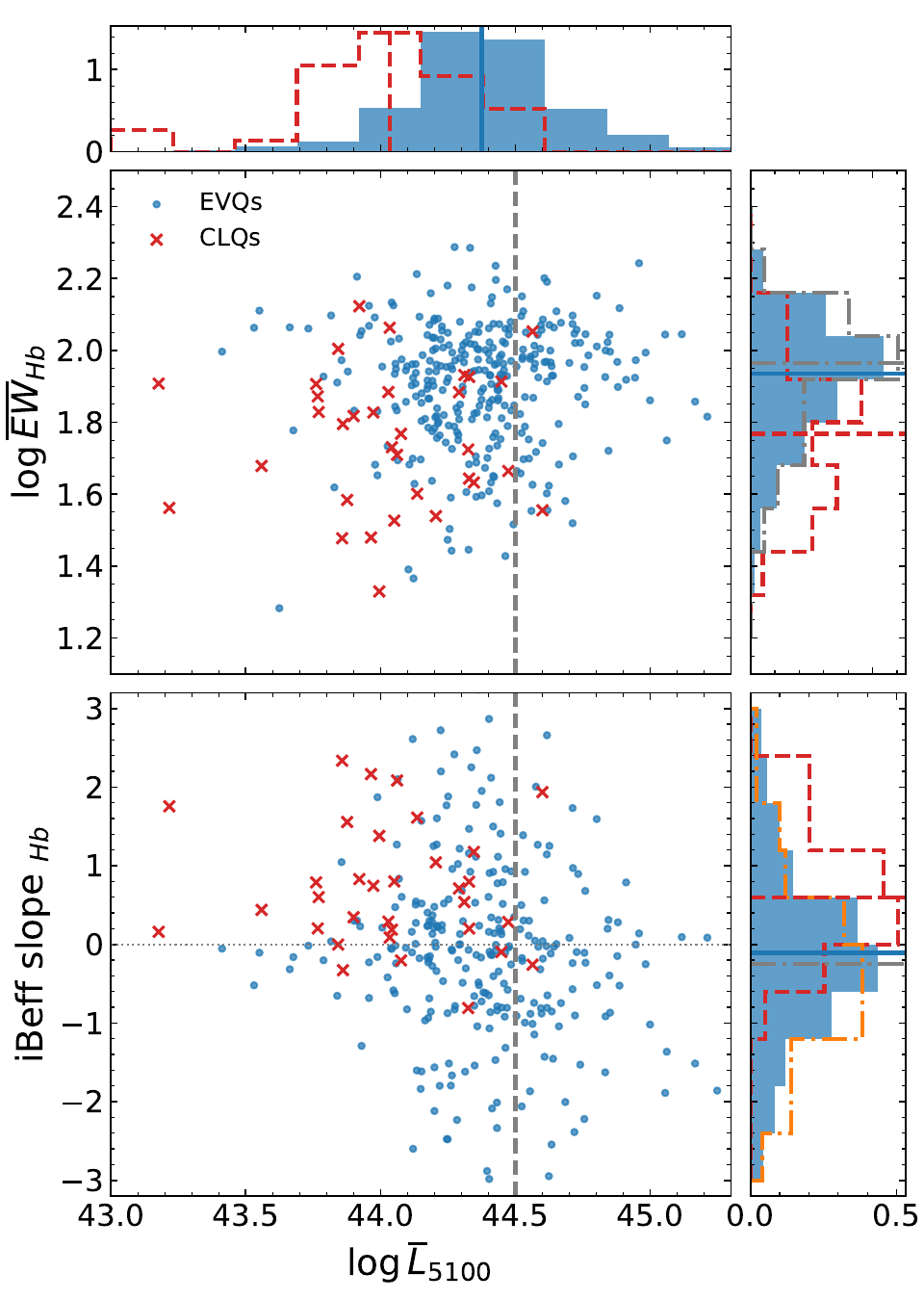}
   \caption{The distribution of \Hb\ EW and iBeff slope  versus the mean continuum luminosity. The iBeff slope is given by the $\Delta EW_{Hb}$/$\Delta L_{5100}$ between the brightest and dimmest states. The histogram distributions of the parameters are shown aside the scatter plot. The median of each distribution is shown with vertical/horizon lines in accordance style. From the EVQ sample, the luminous sub-sample with $\log \overline{L}_{5100}>44.5$ are plotted with gray lines in the histograms.
   \label{fig:CLQ_iBeff}}
\end{figure}
\begin{figure}[tb!]
   \includegraphics[width=0.48\textwidth]{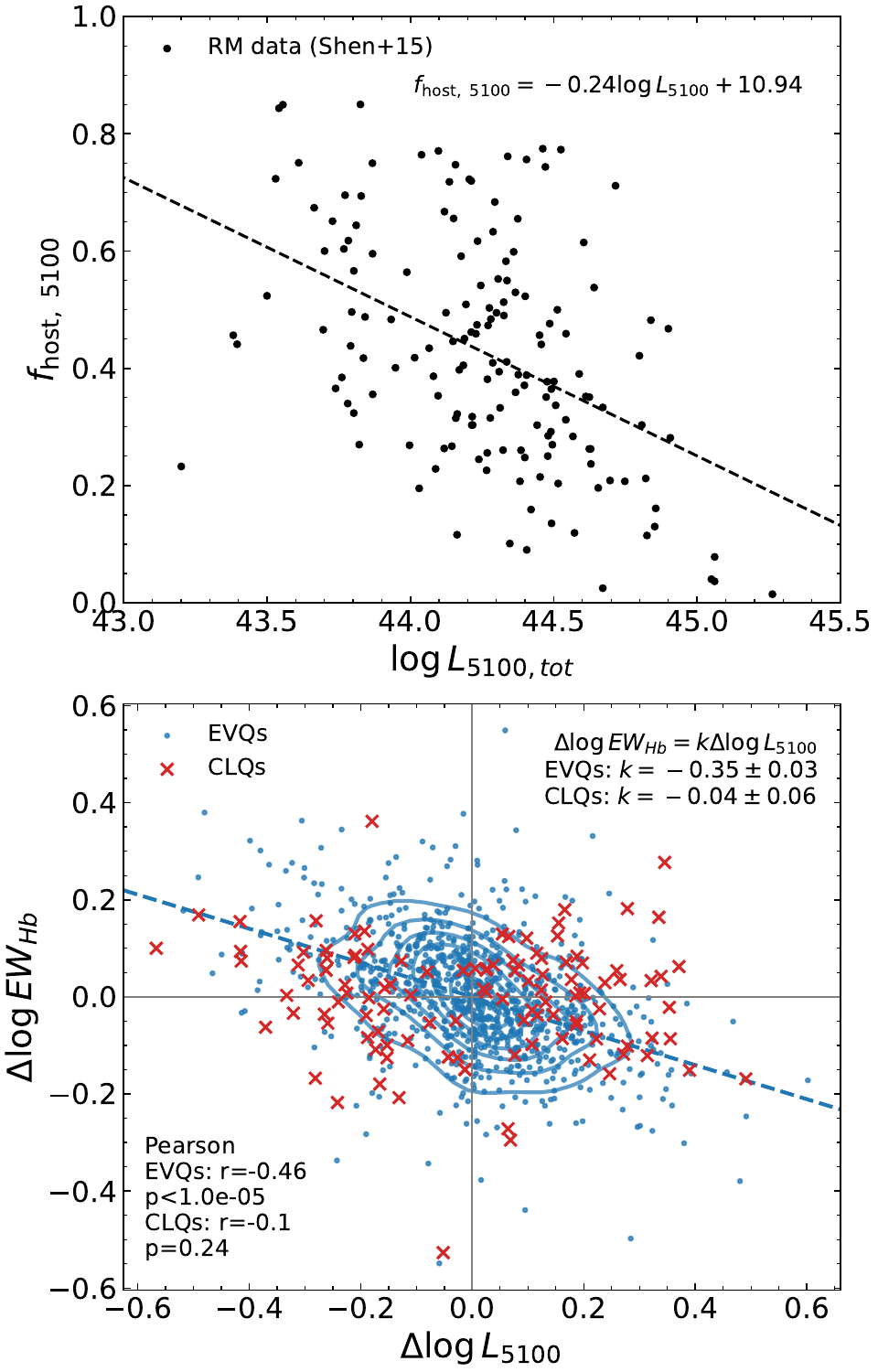}
   \caption{Upper: the correlation between the total luminosity at 5100$\AA$ and the host galaxy fraction from the RM spectra of \citet{Shen2015}. The dashed line represents a simple linear regression (adopting $L_{5100}$ as the independent variable), which is then utilized to estimate the host fraction at given input $L_{5100}$ for our EVQs/CLQs plotted in the lower panel. Lower: the iBeff of \Hb\ of the EVQs and CLQs with the host contamination corrected. The orthogonal distance regression fitting results and the Spearman correlation coefficient $r$ (with p-values) of each sample are listed.
   \label{fig:host_corr}}
\end{figure}

In Fig. \ref{fig:CLQ_EW}, we plot the variation of line EW along with the continuum of our \Hb\ EVQ sample, in comparison with the 33 archived \Hb-selected CLQs. As already shown in Fig. \ref{fig:Delta_L}, there is no clear correlation between $\Delta EW_{Hb}$ and $\Delta L_{5100}$ in the EVQ sample, i.e., no iBeff. Meanwhile a prominent positive correlation is observed in the CLQ sample (i.e., a reverse iBeff). This is actually expected as the definition of changing-look automatically requires much weaker \Hb\ broad line in dim-state CLQs. However, when it comes to the \MgII\ line for the same \Hb-selected CLQs (i.e., lower panel of Fig. \ref{fig:CLQ_EW}), CLQs also exhibit iBeff (suggesting CLQs are physically similar to normal quasars), but with a slope marginally flatter than EVQs.

We further examine the mean broad \Hb\ EW versus the mean 5100\AA\ luminosity, and \Hb\ iBeff slope versus the mean 5100\AA\ luminosity, for the EVQs and \Hb-selected CLQs (Fig. \ref{fig:CLQ_iBeff}). The iBeff slope is calculated using the brightest and dimmest epochs of each source. Similar to the results in Fig. \ref{fig:CLQ_EW}, the majority of CLQs have a reverse iBeff slope, with a median of $0.60\pm0.18$
\footnote{We note that the only study in literature discussing the change in EW of CLQs is presented in \cite{Green2022}, which reports a notably higher average iBeff slope of 2.05 compared to ours. However, their result could be severely influenced by an individual target J091234.00+262828.32, which has an iBeff slope of $\sim$107 due to its minor change in $L_{5100}$, primarily affected by the host galaxy. The average slope of the rest CLQs of \cite{Green2022} is comparable with our result.}
, while the median for EVQs is $-0.11\pm0.05$. The overall luminosity and mean \Hb\ EW of CLQs are lower than those of EVQs. However we do not see clear correlation between mean broad \Hb\ EW and 5100\AA\ luminosity, or between iBeff slope and 5100\AA\ luminosity, for our uniformly selected EVQ sample. The smaller mean \Hb\ EW and 5100\AA\ luminosity of CLQs (compared with EVQs) are likely caused by observational biases, that EVQs with weaker broad \Hb\ line and lower 5100\AA\ luminosity are more likely be classified as CLQs (due to limited spectral SNR). 

Could the different iBeff slope of CLQs (compared with EVQs, as shown in Fig. \ref{fig:CLQ_EW} and \ref{fig:CLQ_iBeff}) be due to potentially stronger host contamination in CLQs? Given the overall low SNR of the SDSS spectra, directly decomposing the host component in individual sources through spectral fitting is yet unreliable. Nevertheless, by referencing to the results from \citet{Shen2015}, we can estimate the host fraction as a simple function of the luminosity at 5100\AA\ and give a rough correction for host contamination. In the upper panel of Fig. \ref{fig:host_corr}, we present the correlation between the total luminosity (AGN plus host) at 5100\AA, $L_{5100, tot}$, and the host fraction at 5100\AA, $f_{host, 5100}$, decomposed by \citet{Shen2015} utilizing their reverberation mapping data. Through simple linear regression we yield the expected average $f_{host, 5100}$ as a function of $L_{5100, tot}$. For each EVQ/CLQ, we conservatively estimate its host contribution using the $L_{5100, tot}$ from its dimmest epoch and subtract the host contribution from the spectra of all epochs. The corrected result is shown in the lower panel of Fig. \ref{fig:host_corr}, in which we could see clear \Hb\ iBeff in our EVQs, with a slope similar to that from the luminous sub-sample (Fig. \ref{fig:HighL_iBeff}) and closer to the results in literature \citep{Goad2004,Rakic2017}. This further suggests the absence of \Hb\ iBeff in EVQs shown in Fig. \ref{fig:Delta_L} is due to host contamination. However, after the correction, the average iBeff slope of CLQs remains statistically different from that of EVQs, which could be an expected consequence of the definition of ``changing-look".

\subsubsection{Are CLQs a distinct population or just a biased sub-population of EVQs?}\label{subsec:biasedpopulation}

As we have shown above, the initial definition of ``changing-look" requires the disappearance of broad \Hb\ line in the spectrum of dim state, but note the definition is qualitative but not quantitative. In reality, the broad \Hb\ lines  of CLQs selected in literature do weaken in the dim state, but not ``completely disappear" in most of them.

Meanwhile, the broad \Hb\ line EW of our EVQs does not systematically vary with continuum, i.e., on average with no iBeff, but with large scatter in the iBeff slope in individual sources. Thus individual EVQs with reverse iBeff slope (smaller broad \Hb\ line EW in dimmer state) could more likely be identified as ``changing-look" due to the limited SNR of the spectra of dim state. Meanwhile the CLQs selected in literature do exhibit reverse \Hb\ iBeff as expected.  

In this sense, CLQs are not necessarily a distinct population, but a sub-population (biased by its definition) of EVQs. In this scheme, the fundamental driven mechanism for both CLQs and EVQs is the continuum variation. 

Previous researches implied that the lower Eddington ratio should be responsible for the dramatic change of the BELs in CLQs \citep[e.g.][]{MacLeod2016}. Similarly, as shown in Fig. \ref{fig:CLQ_iBeff}, we show CLQs have systematically lower luminosity and lower \Hb\ EW, compared with our EVQs. However, all these facts could be attributed to selection biases: 1) the stronger host contamination in sources with lower luminosity/Eddington ratio could hinder the detection of broad \Hb\ in the noisy dim state spectra; 2) intrinsically weaker \Hb\ lines are more difficult to be detected in the noisy dim state spectra.

We note in literature while most CLQs were selected based on \Hb\ line, \MgII\ or \CIV\ CLQs were rarely found \citep{Guo2020a, Ross2020}. This could also be naturally be attributed to the different iBeff slopes of \Hb, \MgII\ and \CIV\ we shown in Fig. \ref{fig:Delta_L}. Since both \MgII\ and \CIV\ of EVQs exhibit strong iBeff, sources in the dim state tend to have larger EW of \MgII\ and \CIV, thus they are less likely be classified as changing-look. Meanwhile, as EVQs exhibit no apparent iBeff on average in \Hb\ line (like due to host contamination), because of the large scatter in the iBeff slope from source to source, some EVQs could exhibit reverse iBeff in \Hb\ and are more likely to identified as changing-look in their dim states. 

The large scatter in the observed iBeff slope may partially be attributed to the lag between broad line and continuum variation. Meanwhile, the fact that the observed optical continuum variation does not necessarily coordinate with that of the UV ionizing continuum \citep{Xin2020,Gaskell2021} could also play a significant role. Note such non-coordinated variations in different bands (see also \citealt{Sou2022}) are natural subsequence of the inhomogeneous disc fluctuation model \citep{Cai2018a,Cai2020}.

To conclude, though we can not completely rule out dramatic vanishment of BLR in some CLQs, most CLQs selected in literature appear consistent with a biased sub-population of normal quasars with extreme continuum variability, i.e.,  the essence of the ``changing-look" phenomena is the extreme continuum variability.


\subsection{Line Breathing} \label{subsec:breathing}

As shown in Fig. \ref{fig:Delta_L}, the breathing of \Hb\ is missing in our EVQ sample, contradicting the results of previous works on AGNs and quasars \citep[e.g., ][]{Denney2009, Park2012, Wang2020}. In addition, the breathing relation in the Hb line within CLQ sample is also insignificant, with a Spearman test $r=-0.29$ and $p=1.0\times10^{-2}$.

It is not uncommon for AGNs to be caught undergoing changes in their breathing relations. For example, \citet{Lu2022} reveal that the \Hb\ of NGC 5548 well breathes in the most recent 5 years, however, previous decades of data deviate significantly. Another interesting example is Mrk 50, in which a breathing mode transition (from anti-breathing to normal breathing) was captured during an $\sim$ 80-day RM observation (cf. Fig. 16 and Fig. 21 in \citealt{Barth2015a}). A more interesting fact is that the breathing period of NGC 5548 coincides with its luminous stage, while Mrk 50 restored its breathing when it was at its dimmest state, thus it seems the breathing mode transition may not be simply one-way related to the flux level. Furthermore, although the majority of sources in \citet{Wang2020} exhibit \Hb\ breathing, a considerable proportion of their sources exhibit anti-breathing. This suggests that breathing mode transitions might be frequent and the structure of the BLRs could be more complex than expected in the simple breathing model.

Meanwhile, it is notable that more than half of the sources in our EVQ sample have a maximum spectral gap $>$3000 days which is much longer than the duration of common RM projects. Therefore the absence of significant breathing/anti-breathing on average in \Hb\ (and in \MgII, and marginal in \CIV) in our EVQs could suggest possibly different evolution of the BLR (i.e., no breathing on average) on longer timescales. 


\section{Conclusions} \label{sec:conclusion}

In this work, we measured all 20,069 available SDSS spectra for the 14,012 EVQs built in \citetalias{Ren2022} with {\tt PyQSOFit}. The catalog is presented in \S\ref{B_catalog}. From this sample, we select 1259 EVQs with multi-epoch SDSS spectra after eliminating spectra with suspicious calibrations, and study the spectral variations in individual EVQs.  

We find clear "bluer-when-brighter" relation in multi-epoch spectra of EVQs, consistent with previous works on normal quasars and AGNs. The iBeff of the broad \MgII\ and \CIV\ in EVQs are significantly seen. However, no robust iBeff of \Hb\ is detected, which could be attributed to stronger host contamination at longer rest-frame wavelengths. Meanwhile, no systematical variation of the broad line shape and asymmetry with continuum flux is found. We do not detect statistically significant broad line breathing of \Hb, \MgII\ or \CIV\ either, suggesting the BLR evolution on longer timescales ($\sim$ 3000 days in the observed frame) could be different from those on shorter timescales.

More interestingly, through comparing the iBeff of our EVQs with CLQs selected in literature, we show that CLQs are more likely a biased (because of its definition) sub-population of normal quasars with extreme continuum variation, instead of a distinct population. 


\begin{acknowledgments}
The work is supported by National Natural Science Foundation of China (grants No. 11890693, 12033006, \& 12192221), and the Cyrus Chung Ying Tang Foundations.

\end{acknowledgments}

\vspace{5mm}
\facilities{Sloan, PS1, PO:1.2m}

\software{PyQSOFit \citep{2018ascl.soft09008G},
          astropy \citep{astropy:2013, astropy:2018, astropy:2022},
          pandas \citep{reback2020pandas, mckinney-proc-scipy-2010},
          scipy \citep{2020SciPy-NMeth},
          numpy \citep{harris2020array},
          }

\appendix

\section{The EVQ Catalog} \label{B_catalog}

We publish the EVQs catalog along with all measured spectral quantities in this paper. The format of the catalog is described in Table \ref{tab:catalog} and the full catalog is available at \url{https://doi.org/10.5281/zenodo.8328175}.
 
\startlongtable
\centerwidetable
\begin{deluxetable*}{rlccp{0.45\linewidth}}
\tablecaption{FITS catalog format}\label{tab:catalog}
\tablehead{
    \colhead{Number} & \colhead{Column Name} & \colhead{Format} & \colhead{Unit} & \colhead{Description}
}
\decimalcolnumbers
\startdata
        0 & SDSS\_NAME             & string      &                    & Unique identifier from the SDSS DR14 quasar catalog                                                                                      \\
        1 & RA                    & float32     & deg                & Right ascension (J2000)                                                                                                                  \\
        2 & DEC                   & float32     & deg                & Declination (J2000)                                                                                                                      \\
        3 & Z                     & float32     &                    & Redshift                                                                                                                                 \\
        4 & MEANMAG\_G             & float32     & mag                & Weighted mean magnitude of g-band light curve                                                                                            \\
        5 & NG                    & int32       &                    & Number of observations in g-band light curve                                                                                             \\
        6 & PLATE                 & int32       &                    & SDSS plate number                                                                                                                        \\
        7 & MJD                   & int32       &                    & MJD when spectrum was observed                                                                                                           \\
        8 & FIBER                 & int32       &                    & SDSS fiber ID                                                                                                                            \\
        9 & SPECPRIMARY           & bool        &                    & If the spectrum is the primary observation of object                                                                                     \\
       10 & SPECMAG\_G             & float32     & mag                & Spectrum projected onto g filters                                                                                                        \\
       11 & SPECMAG\_G\_ERR         & float32     & mag                & Error in SPECMAG\_G                                                                                                                       \\
       12 & SPECMAG\_R             & float32     & mag                & Spectrum projected onto r filters                                                                                                        \\
       13 & SPECMAG\_R\_ERR         & float32     & mag                & Error in SPECMAG\_R                                                                                                                       \\
       14 & SN\_CONTI              & float32     &                    & Mean signal-to-noise ratio per pixel of continuum estimated at wavelength around 1350, 3000, and 5100 depending on the spectral coverage \\
       15 & PL\_SLOPE              & float32     &                    & Slope of AGN power law                                                                                                                   \\
       16 & PL\_SLOPE\_ERR          & float32     &                    & Error in PL\_SLOPE                                                                                                                        \\
       17 & LOGL5100              & float32     & $\ergs$            & Logarithmic continuum luminosity at rest-frame 5100$\gaa$                                                                             \\
       18 & LOGL5100\_ERR          & float32     & $\ergs$            & Error in LOGL5100                                                                                                                        \\
       19 & LOGL3000              & float32     & $\ergs$            & Logarithmic continuum luminosity at rest-frame 3000$\gaa$                                                                             \\
       20 & LOGL3000\_ERR          & float32     & $\ergs$            & Error in LOGL3000                                                                                                                        \\
       21 & LOGL1350              & float32     & $\ergs$            & Logarithmic continuum luminosity at rest-frame 1350$\gaa$                                                                             \\
       22 & LOGL1350\_ERR          & float32     & $\ergs$            & Error in LOGL1350                                                                                                                        \\
       23 & FE\_2240\_2650\_FLUX     & float32     & $10^{-17}\flux$ & Rest-frame flux of the UV Fe II complex within the 2240-2650$\gaa$                                                                    \\
       24 & FE\_2240\_2650\_FLUX\_ERR & float32     & $10^{-17}\flux$ & Error in FE\_2240\_2650\_FLUX                                                                                                               \\
       25 & FE\_2240\_2650\_EW       & float32     & $\gaa$           & Rest-frame EW of the UV Fe II complex within the 2240-2650$\gaa$                                                                      \\
       26 & FE\_2240\_2650\_EW\_ERR   & float32     & $\gaa$           & Error in FE\_2240\_2650\_EW                                                                                                                 \\
       27 & FE\_4435\_4685\_FLUX     & float32     & $10^{-17}\flux$ & Rest-frame flux of the optical Fe II complex within the 4435-4685$\gaa$                                                               \\
       28 & FE\_4435\_4685\_FLUX\_ERR & float32     & $10^{-17}\flux$ & Error in FE\_4435\_4685\_FLUX                                                                                                               \\
       29 & FE\_4435\_4685\_EW       & float32     & $\gaa$           & Rest-frame EW of the optical Fe II complex within the 4435\_4685$\gaa$                                                                 \\
       30 & FE\_4435\_4685\_EW\_ERR   & float32     & $\gaa$           & Error in FE\_4435\_4685\_EW                                                                                                                 \\
       31 & HB\_BR\_PEAK            & float32     & $\gaa$           & Peak wavelength of Hb broad component                                                                                                    \\
       32 & HB\_BR\_PEAK\_ERR        & float32     & $\gaa$           & Error in HB\_BR\_PEAK                                                                                                                      \\
       33 & HB\_BR\_BISECT          & float32     & $\gaa$           & Wavelength bisect the area of Hb broad component                                                                                         \\
       34 & HB\_BR\_BISECT\_ERR      & float32     & $\gaa$           & Error in HB\_BR\_BISECT                                                                                                                    \\
       35 & HB\_BR\_FLUX            & float32     & $10^{-17}\flux$ & Flux of Hb broad component                                                                                                               \\
       36 & HB\_BR\_FLUX\_ERR        & float32     & $10^{-17}\flux$ & Error in HB\_BR\_FLUX                                                                                                                      \\
       37 & HB\_BR\_EW              & float32     & $\gaa$           & Rest-frame EW of Hb broad component                                                                                                      \\
       38 & HB\_BR\_EW\_ERR          & float32     & $\gaa$           & Error in HB\_BR\_EW                                                                                                                        \\
       39 & HB\_BR\_SIGMA           & float32     & $\kms$             & Line dispersion of Hb broad component                                                                                                    \\
       40 & HB\_BR\_SIGMA\_ERR       & float32     & $\kms$             & Error in HB\_BR\_SIGMA                                                                                                                     \\
       41 & HB\_BR\_FWHM            & float32     & $\kms$             & FWHM of Hb broad component                                                                                                               \\
       42 & HB\_BR\_FWHM\_ERR        & float32     & $\kms$             & Error in HB\_BR\_FWHM                                                                                                                      \\
       43 & HB\_BR\_FWQM            & float32     & $\kms$             & FWQM of Hb broad component                                                                                                               \\
       44 & HB\_BR\_FWQM\_ERR        & float32     & $\kms$             & Error in HB\_BR\_FWQM                                                                                                                      \\
       45 & HB\_BR\_FW10M           & float32     & $\kms$             & FW10M of Hb broad component                                                                                                              \\
       46 & HB\_BR\_FW10M\_ERR       & float32     & $\kms$             & Error in HB\_BR\_FW10M                                                                                                                     \\
       47 & HB\_BR\_Z50             & float32     & $\kms$             & The half maximum center shift of Hb broad component                                                                                      \\
       48 & HB\_BR\_Z50\_ERR         & float32     & $\kms$             & Error in HB\_BR\_Z50                                                                                                                       \\
       49 & HB\_BR\_Z25             & float32     & $\kms$             & The quarter maximum center shift of Hb broad component                                                                                   \\
       50 & HB\_BR\_Z25\_ERR         & float32     & $\kms$             & Error in HB\_BR\_Z25                                                                                                                       \\
       51 & HB\_BR\_Z10             & float32     & $\kms$             & The 10 percent maximum center shift of Hb broad component                                                                                \\
       52 & HB\_BR\_Z10\_ERR         & float32     & $\kms$             & Error in HB\_BR\_Z10                                                                                                                       \\
       53 & HB\_NA\_PEAK            & float32     & $\gaa$           & Peak wavelength of Hb narrow component                                                                                                   \\
       54 & HB\_NA\_PEAK\_ERR        & float32     & $\gaa$           & Error in HB\_NA\_PEAK                                                                                                                      \\
       55 & HB\_NA\_FLUX            & float32     & $10^{-17}\flux$ & Flux of Hb narrow component                                                                                                              \\
       56 & HB\_NA\_FLUX\_ERR        & float32     & $10^{-17}\flux$ & Error in HB\_NA\_FLUX                                                                                                                      \\
       57 & HB\_NA\_EW              & float32     & $\gaa$           & Rest-frame EW of Hb narrow component                                                                                                     \\
       58 & HB\_NA\_EW\_ERR          & float32     & $\gaa$           & Error in HB\_NA\_EW                                                                                                                        \\
       59 & HB\_NA\_SIGMA           & float32     & $\kms$             & Line dispersion of Hb narrow component                                                                                                   \\
       60 & HB\_NA\_SIGMA\_ERR       & float32     & $\kms$             & Error in HB\_NA\_SIGMA                                                                                                                     \\
       61 & HB\_NA\_FWHM            & float32     & $\kms$             & FWHM of Hb narrow component                                                                                                              \\
       62 & HB\_NA\_FWHM\_ERR        & float32     & $\kms$             & Error in HB\_NA\_FWHM                                                                                                                      \\
       63 & OIII4959C\_PEAK        & float32     & $\gaa$           & Peak wavelength of OIII4959 core component                                                                                               \\
       64 & OIII4959C\_PEAK\_ERR    & float32     & $\gaa$           & Error in OIII4959C\_PEAK                                                                                                                  \\
       65 & OIII4959C\_FLUX        & float32     & $10^{-17}\flux$ & Flux of OIII4959 core component                                                                                                          \\
       66 & OIII4959C\_FLUX\_ERR    & float32     & $10^{-17}\flux$ & Error in OIII4959C\_FLUX                                                                                                                  \\
       67 & OIII4959C\_EW          & float32     & $\gaa$           & Rest-frame EW of OIII4959 core component                                                                                                 \\
       68 & OIII4959C\_EW\_ERR      & float32     & $\gaa$           & Error in OIII4959C\_EW                                                                                                                    \\
       69 & OIII4959C\_SIGMA       & float32     & $\kms$             & Line dispersion of OIII4959 core component                                                                                               \\
       70 & OIII4959C\_SIGMA\_ERR   & float32     & $\kms$             & Error in OIII4959C\_SIGMA                                                                                                                 \\
       71 & OIII4959C\_FWHM        & float32     & $\kms$             & FWHM of OIII4959 core component                                                                                                          \\
       72 & OIII4959C\_FWHM\_ERR    & float32     & $\kms$             & Error in OIII4959C\_FWHM                                                                                                                  \\
       73 & OIII4959W\_PEAK        & float32     & $\gaa$           & Peak wavelength of OIII4959 wing component                                                                                               \\
       74 & OIII4959W\_PEAK\_ERR    & float32     & $\gaa$           & Error in OIII4959W\_PEAK                                                                                                                  \\
       75 & OIII4959W\_FLUX        & float32     & $10^{-17}\flux$ & Flux of OIII4959 wing component                                                                                                          \\
       76 & OIII4959W\_FLUX\_ERR    & float32     & $10^{-17}\flux$ & Error in OIII4959W\_FLUX                                                                                                                  \\
       77 & OIII4959W\_EW          & float32     & $\gaa$           & Rest-frame EW of OIII4959 wing component                                                                                                 \\
       78 & OIII4959W\_EW\_ERR      & float32     & $\gaa$           & Error in OIII4959W\_EW                                                                                                                    \\
       79 & OIII4959W\_SIGMA       & float32     & $\kms$             & Line dispersion of OIII4959 wing component                                                                                               \\
       80 & OIII4959W\_SIGMA\_ERR   & float32     & $\kms$             & Error in OIII4959W\_SIGMA                                                                                                                 \\
       81 & OIII4959W\_FWHM        & float32     & $\kms$             & FWHM of OIII4959 wing component                                                                                                          \\
       82 & OIII4959W\_FWHM\_ERR    & float32     & $\kms$             & Error in OIII4959W\_FWHM                                                                                                                  \\
       83 & OIII5007C\_PEAK        & float32     & $\gaa$           & Peak wavelength of OIII5007 core component                                                                                               \\
       84 & OIII5007C\_PEAK\_ERR    & float32     & $\gaa$           & Error in OIII5007C\_PEAK                                                                                                                  \\
       85 & OIII5007C\_FLUX        & float32     & $10^{-17}\flux$ & Flux of OIII5007 core component                                                                                                          \\
       86 & OIII5007C\_FLUX\_ERR    & float32     & $10^{-17}\flux$ & Error in OIII5007C\_FLUX                                                                                                                  \\
       87 & OIII5007C\_EW          & float32     & $\gaa$           & Rest-frame EW of OIII5007 core component                                                                                                 \\
       88 & OIII5007C\_EW\_ERR      & float32     & $\gaa$           & Error in OIII5007C\_EW                                                                                                                    \\
       89 & OIII5007C\_SIGMA       & float32     & $\kms$             & Line dispersion of OIII5007 core component                                                                                               \\
       90 & OIII5007C\_SIGMA\_ERR   & float32     & $\kms$             & Error in OIII5007C\_SIGMA                                                                                                                 \\
       91 & OIII5007C\_FWHM        & float32     & $\kms$             & FWHM of OIII5007 core component                                                                                                          \\
       92 & OIII5007C\_FWHM\_ERR    & float32     & $\kms$             & Error in OIII5007C\_FWHM                                                                                                                  \\
       93 & OIII5007W\_PEAK        & float32     & $\gaa$           & Peak wavelength of OIII5007 wing component                                                                                               \\
       94 & OIII5007W\_PEAK\_ERR    & float32     & $\gaa$           & Error in OIII5007W\_PEAK                                                                                                                  \\
       95 & OIII5007W\_FLUX        & float32     & $10^{-17}\flux$ & Flux of OIII5007 wing component                                                                                                          \\
       96 & OIII5007W\_FLUX\_ERR    & float32     & $10^{-17}\flux$ & Error in OIII5007W\_FLUX                                                                                                                  \\
       97 & OIII5007W\_EW          & float32     & $\gaa$           & Rest-frame EW of OIII5007 wing component                                                                                                 \\
       98 & OIII5007W\_EW\_ERR      & float32     & $\gaa$           & Error in OIII5007W\_EW                                                                                                                    \\
       99 & OIII5007W\_SIGMA       & float32     & $\kms$             & Line dispersion of OIII5007 wing component                                                                                               \\
      100 & OIII5007W\_SIGMA\_ERR   & float32     & $\kms$             & Error in OIII5007W\_SIGMA                                                                                                                 \\
      101 & OIII5007W\_FWHM        & float32     & $\kms$             & FWHM of OIII5007 wing component                                                                                                          \\
      102 & OIII5007W\_FWHM\_ERR    & float32     & $\kms$             & Error in OIII5007W\_FWHM                                                                                                                  \\
      103 & MGII\_BR\_PEAK          & float32     & $\gaa$           & Peak wavelength of MgII broad component                                                                                                  \\
      104 & MGII\_BR\_PEAK\_ERR      & float32     & $\gaa$           & Error in MGII\_BR\_PEAK                                                                                                                    \\
      105 & MGII\_BR\_BISECT        & float32     & $\gaa$           & Wavelength bisect the area of MgII broad component                                                                                       \\
      106 & MGII\_BR\_BISECT\_ERR    & float32     & $\gaa$           & Error in MGII\_BR\_BISECT                                                                                                                  \\
      107 & MGII\_BR\_FLUX          & float32     & $10^{-17}\flux$ & Flux of MgII broad component                                                                                                             \\
      108 & MGII\_BR\_FLUX\_ERR      & float32     & $10^{-17}\flux$ & Error in MGII\_BR\_FLUX                                                                                                                    \\
      109 & MGII\_BR\_EW            & float32     & $\gaa$           & Rest-frame EW of MgII broad component                                                                                                    \\
      110 & MGII\_BR\_EW\_ERR        & float32     & $\gaa$           & Error in MGII\_BR\_EW                                                                                                                      \\
      111 & MGII\_BR\_SIGMA         & float32     & $\kms$             & Line dispersion of MgII broad component                                                                                                  \\
      112 & MGII\_BR\_SIGMA\_ERR     & float32     & $\kms$             & Error in MGII\_BR\_SIGMA                                                                                                                   \\
      113 & MGII\_BR\_FWHM          & float32     & $\kms$             & FWHM of MgII broad component                                                                                                             \\
      114 & MGII\_BR\_FWHM\_ERR      & float32     & $\kms$             & Error in MGII\_BR\_FWHM                                                                                                                    \\
      115 & MGII\_BR\_FWQM          & float32     & $\kms$             & FWQM of MgII broad component                                                                                                             \\
      116 & MGII\_BR\_FWQM\_ERR      & float32     & $\kms$             & Error in MGII\_BR\_FWQM                                                                                                                    \\
      117 & MGII\_BR\_FW10M         & float32     & $\kms$             & FW10M of MgII broad component                                                                                                            \\
      118 & MGII\_BR\_FW10M\_ERR     & float32     & $\kms$             & Error in MGII\_BR\_FW10M                                                                                                                   \\
      119 & MGII\_BR\_Z50           & float32     & $\kms$             & The half maximum center shift of MgII broad component                                                                                    \\
      120 & MGII\_BR\_Z50\_ERR       & float32     & $\kms$             & Error in MGII\_BR\_Z50                                                                                                                     \\
      121 & MGII\_BR\_Z25           & float32     & $\kms$             & The quarter maximum center shift of MgII broad component                                                                                 \\
      122 & MGII\_BR\_Z25\_ERR       & float32     & $\kms$             & Error in MGII\_BR\_Z25                                                                                                                     \\
      123 & MGII\_BR\_Z10           & float32     & $\kms$             & The 10 percent maximum center shift of MgII broad component                                                                              \\
      124 & MGII\_BR\_Z10\_ERR       & float32     & $\kms$             & Error in MGII\_BR\_Z10                                                                                                                     \\
      125 & MGII\_NA\_PEAK          & float32     & $\gaa$           & Peak wavelength of MgII narrow component                                                                                                 \\
      126 & MGII\_NA\_PEAK\_ERR      & float32     & $\gaa$           & Error in MGII\_NA\_PEAK                                                                                                                    \\
      127 & MGII\_NA\_FLUX          & float32     & $10^{-17}\flux$ & Flux of MgII narrow component                                                                                                            \\
      128 & MGII\_NA\_FLUX\_ERR      & float32     & $10^{-17}\flux$ & Error in MGII\_NA\_FLUX                                                                                                                    \\
      129 & MGII\_NA\_EW            & float32     & $\gaa$           & Rest-frame EW of MgII narrow component                                                                                                   \\
      130 & MGII\_NA\_EW\_ERR        & float32     & $\gaa$           & Error in MGII\_NA\_EW                                                                                                                      \\
      131 & MGII\_NA\_SIGMA         & float32     & $\kms$             & Line dispersion of MgII narrow component                                                                                                 \\
      132 & MGII\_NA\_SIGMA\_ERR     & float32     & $\kms$             & Error in MGII\_NA\_SIGMA                                                                                                                   \\
      133 & MGII\_NA\_FWHM          & float32     & $\kms$             & FWHM of MgII narrow component                                                                                                            \\
      134 & MGII\_NA\_FWHM\_ERR      & float32     & $\kms$             & Error in MGII\_NA\_FWHM                                                                                                                    \\
      135 & CIV\_BR\_PEAK           & float32     & $\gaa$           & Peak wavelength of CIV broad component                                                                                                   \\
      136 & CIV\_BR\_PEAK\_ERR       & float32     & $\gaa$           & Error in CIV\_BR\_PEAK                                                                                                                     \\
      137 & CIV\_BR\_BISECT         & float32     & $\gaa$           & Wavelength bisect the area of CIV broad component                                                                                        \\
      138 & CIV\_BR\_BISECT\_ERR     & float32     & $\gaa$           & Error in CIV\_BR\_BISECT                                                                                                                   \\
      139 & CIV\_BR\_FLUX           & float32     & $10^{-17}\flux$ & Flux of CIV broad component                                                                                                              \\
      140 & CIV\_BR\_FLUX\_ERR       & float32     & $10^{-17}\flux$ & Error in CIV\_BR\_FLUX                                                                                                                     \\
      141 & CIV\_BR\_EW             & float32     & $\gaa$           & Rest-frame EW of CIV broad component                                                                                                     \\
      142 & CIV\_BR\_EW\_ERR         & float32     & $\gaa$           & Error in CIV\_BR\_EW                                                                                                                       \\
      143 & CIV\_BR\_SIGMA          & float32     & $\kms$             & Line dispersion of CIV broad component                                                                                                   \\
      144 & CIV\_BR\_SIGMA\_ERR      & float32     & $\kms$             & Error in CIV\_BR\_SIGMA                                                                                                                    \\
      145 & CIV\_BR\_FWHM           & float32     & $\kms$             & FWHM of CIV broad component                                                                                                              \\
      146 & CIV\_BR\_FWHM\_ERR       & float32     & $\kms$             & Error in CIV\_BR\_FWHM                                                                                                                     \\
      147 & CIV\_BR\_FWQM           & float32     & $\kms$             & FWQM of CIV broad component                                                                                                              \\
      148 & CIV\_BR\_FWQM\_ERR       & float32     & $\kms$             & Error in CIV\_BR\_FWQM                                                                                                                     \\
      149 & CIV\_BR\_FW10M          & float32     & $\kms$             & FW10M of CIV broad component                                                                                                             \\
      150 & CIV\_BR\_FW10M\_ERR      & float32     & $\kms$             & Error in CIV\_BR\_FW10M                                                                                                                    \\
      151 & CIV\_BR\_Z50            & float32     & $\kms$             & The half maximum center shift of CIV broad component                                                                                     \\
      152 & CIV\_BR\_Z50\_ERR        & float32     & $\kms$             & Error in CIV\_BR\_Z50                                                                                                                      \\
      153 & CIV\_BR\_Z25            & float32     & $\kms$             & The quarter maximum center shift of CIV broad component                                                                                  \\
      154 & CIV\_BR\_Z25\_ERR        & float32     & $\kms$             & Error in CIV\_BR\_Z25                                                                                                                      \\
      155 & CIV\_BR\_Z10            & float32     & $\kms$             & The 10 percent maximum center shift of CIV broad component                                                                               \\
      156 & CIV\_BR\_Z10\_ERR        & float32     & $\kms$             & Error in CIV\_BR\_Z10                                                                                                                      \\
      157 & LOGLBOL\_5100          & float32     & $\ergs$            & Logarithmic bolometric luminosity estimated based on LOGL5100                                                                            \\
      158 & LOGLBOL\_5100\_ERR      & float32     & $\ergs$            & Error in LOGLBOL\_5100                                                                                                                    \\
      159 & LOGLBOL\_3000          & float32     & $\ergs$            & Logarithmic bolometric luminosity estimated based on LOGL3000                                                                            \\
      160 & LOGLBOL\_3000\_ERR      & float32     & $\ergs$            & Error in LOGLBOL\_3000                                                                                                                    \\
      161 & LOGLBOL\_1350          & float32     & $\ergs$            & Logarithmic bolometric luminosity estimated based on LOGL1350                                                                            \\
      162 & LOGLBOL\_1350\_ERR      & float32     & $\ergs$            & Error in LOGLBOL\_1350                                                                                                                    \\
      163 & LOGMBH\_HB             & float32     & $M_\odot$              & Logarithmic black hole mass estimated based on broad Hb line                                                                             \\
      164 & LOGMBH\_HB\_ERR         & float32     & $M_\odot$              & Error in LOGMBH\_HB                                                                                                                       \\
      165 & LOGMBH\_MGII           & float32     & $M_\odot$              & Logarithmic black hole mass estimated based on broad MgII line                                                                           \\
      166 & LOGMBH\_MGII\_ERR       & float32     & $M_\odot$              & Error in LOGMBH\_MGII                                                                                                                     \\
      167 & LOGMBH\_CIV            & float32     & $M_\odot$              & Logarithmic black hole mass estimated based on broad CIV line                                                                            \\
      168 & LOGMBH\_CIV\_ERR        & float32     & $M_\odot$              & Error in LOGMBH\_CIV                                                                                                                      \\
      169 & LOGLBOL               & float32     & $\ergs$            & The adopted fiducial bolometric luminosity                                                                                               \\
      170 & LOGLBOL\_ERR           & float32     & $\ergs$            & Error in LOGLBOL                                                                                                                         \\
      171 & LOGMBH                & float32     & $M_\odot$              & The adopted fiducial black hole mass                                                                                                     \\
      172 & LOGMBH\_ERR            & float32     & $M_\odot$              & Error in LOGMBH                                                                                                                          \\
      173 & LOGREDD               & float32     &                    & Logarithmic Eddington ratio based on fiducial BH mass and bolometric luminosity                                                          \\
\enddata
\tablecomments{We provide all 20,069 available spectral measurements of 14,012 EVQs selected in \citealt{Ren2022}. Repeated observed spectra of a same EVQ will have the same {\tt SDSS\_NAME} with different spectral info ({\tt PLATE},{\tt MJD}, and {\tt FIBER}). We include the {\tt SPECPRIMARY} flag to indicate if the spectrum is the best observation of this object. The unmeasurable parameters are set to -999. The errors are obtained from 100 iterations of Monte Carlo simulation. The complete table is available on \url{https://doi.org/10.5281/zenodo.8328175}}
\end{deluxetable*}

\section{Evaluate spectral calibration} \label{A_drw}

The fiber-drop \citep{Dawson2013} is an unavoidable problem in SDSS spectrophotometry leading to varying degrees of underestimation of spectral flux and has been widely reported in literature \citep[e.g. ][]{Shen2015, Sun2015, Guo2020a}. In certain cases the spectrophotometry could also be overestimated due to contamination to the line of sight by spurious or unrelated signals. Such effects may produce artificial spectral variability which is hard to distinguish from the intrinsic spectral variation. This would be particularly more relevant if one aims to search for rare events (such as CLQs) out of a huge number of spectra. In this work our EVQs are pre-selected based on multi-epoch photometric observations. Below we develop a technique to evaluate the reliability of the SDSS spectrophotometry of the EVQs through comparing the spectroscopic photometry with the photometric light curve modeled with damped random walk (DRW). The spectroscopic photometry which severely deviates from the DRW yielded ranges would be excluded as potentially unreliable.

\subsection{Extend the light curves}

The light curves we built for EVQ selection, consisted of data from SDSS and PS1, only cover a span of 1998 to 2014. About half of the SDSS spectra were conducted by \SDSSIV\ since 2014 \citep{Blanton2017}, significantly beyond the time span of our photometric light curves. We then introduce the PTF/iPTF and ZTF observations to extend the photometric light curves to cover the epochs when \SDSSIV\ spectra were obtained.

The Palomar Transient Factory (PTF) is a fully-automated, wide-field survey conducted by the Palomar 48-inch Samuel Oschin Schmidt telescope with 12K$\times$8K CCD array during 2009 to 2012 \citep{Rau2009,Law2009}. The intermediate Palomar Transient Factory (iPTF) is the successor PTF ran from 2013 to 2017 with a relatively higher cadence. The Zwicky Transient Facility (ZTF) is a next-generation optical time-domain survey build upon the PTF/iPTF. It started from 2018, scanned 3750 $\rm deg^2/hour$, and is still in operation \citep{Masci2018}. The PTF/iPTF and ZTF surveys are shallower than the SDSS and the PS1 with relatively larger photometry error. Nevertheless, among the 3042 EVQs with repeated SDSS spectra, only 11 of them have no detection in either PTF, iPTF or ZTF.
 
\subsection{Evaluating the reliability of spectroscopic photometry}

We use the {\tt celerite} \citep{Foreman-Mackey2017} to model our $g$- and $r$- band light curves and estimate their DRW parameters (timescale $\tau$ and variability amplitude $\sigma$) with maximum likelihood estimation (MLE) method in {\tt scipy}. The mean magnitude $\bar{m}$ of DRW model is assumed to be the mean of the light curve. The detailed explanation of fitting a light curve with DRW process is presented in \citet{Zu2011,Zu2013,Zu2016}.

Next, we utilize the DRW parameters to determine the probability distribution function of the magnitude at a given epoch. According to the DRW process, the next epoch signal ($s_{n+1}$ at $t=t_{n+1}$) is only relative to the present one $s_n$, which can be expressed as:
\begin{equation}
    p(s_{n+1}=x_{n+1}|s_n=x_n) = N(\mu_{n+1},\eta_{n+1}^2),
\end{equation}
where
\begin{equation}
    \mu_{n+1} = e^{-\frac{\Delta t}{\tau}}x_n+\bar{m}(1-e^{-\frac{\Delta t}{\tau}}),
\end{equation}
\begin{equation}
    \eta_{n+1}^2 = \sigma^2(1-e^{-\frac{2\Delta t}{\tau}})+err_n^2+err_{n+1}^2.
\end{equation}
The $s_{n}$ and $s_{n+1}$ are the magnitude of the couple of adjacent light curve data at $t_n$ and $t_{n+1}$ respectively, $err$ the observational error, and $\Delta t=t_{n+1}-t_{n}$.

In any gap of the light curve, with the magnitude known at both ends, we can calculate the probability distribution function (PDF) of the magnitude at any epoch ($t_n<t_0<t_{n+1}$) within this duration. According to Bayes formula, the PDF of the signal at $t_0$ given its adjacent data can be determined as:
\begin{equation}
\begin{split}
    &p(s_0=x_0|s_n=x_n,~s_{n+1}=x_{n+1}) \\
    =&\frac{p(s_0=x_0,~s_{n+1}=x_{n+1}|s_n=x_n)}{p(s_{n+1}=x_{n+1}|s_n=x_n)}\\
    =&\frac{p(s_0=x_0|s_n=x_n)p(s_{n+1}=x_{n+1}|s_0=x_0,~s_n=x_n)}{p(s_{n+1}=x_{n+1}|s_n=x_n)} \\
    =&\frac{p(s_0=x_0|s_n=x_n)p(s_{n+1}=x_{n+1}|s_0=x_0)}{p(s_{n+1}=x_{n+1}|s_n=x_n)}.
\end{split}
\end{equation}

\begin{figure}[tb!]
   \vspace{1em}
   \includegraphics[width=.48\textwidth]{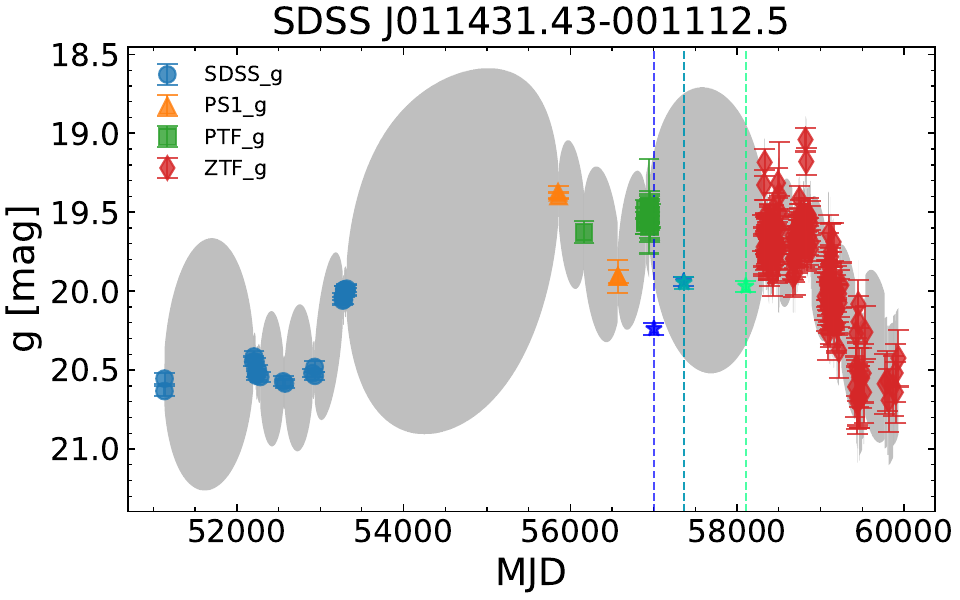}
   \caption{An example of $g$-band light curve showing the 99.7\% confidence intervals of the magnitude in the light curve gaps. The observed photometric magnitudes from different surveys are plotted with distinct colors and marks. The spectroscopic magnitudes of three
   SDSS spectra are marked with stars and vertical dashed lines. As can be seen, the spectroscopic magnitude of the first spectrum is out of the 99.7\% confidence intervals, which is likely due to the fiber-drop effect.
   \label{fig:A1_drwpdf}}
\end{figure}

To evaluate the reliability of a spectroscopic photometry, we first set $t_0$ to the date of the spectrum and find its adjacent photometric data in the corresponding light curve to calculate the PDF of the magnitude at this epoch. By integrating this PDF, we can give the confidence interval of the true magnitude of this spectrum. The sketch of the 99.7\% confidence interval is shown in Fig. \ref{fig:A1_drwpdf}.

In this work, we consider the spectroscopic photometry reliable if both its $g$- and $r$- band spectroscopic magnitude fall within the 99.7\% confidence intervals. Under this criterion, 301 spectra are excluded as potentially unreliable, leaving 1259 sources with repeated observations.

\subsection{The validity of the approach}

Precisely evaluating whether this method can properly rule out unreliable spectroscopic photometry could be challenging. However, we can provide an indirect approach to demonstrate its validity. We further select a sample of low-variability quasars (LVQs) with $|\Delta g|_{\text{max}} < 0.2$ mag, where $|\Delta g|_{\text{max}}$ represents the maximum difference in magnitude between any two epochs in the light curves constructed by \citetalias{Ren2022}. Since we do not expect a sudden change in flux for these low-variability quasars, spectra with significantly deviating flux from their light curve means in LVQs are likely caused by calibration issues.

\begin{figure}[tb!]
   \includegraphics[width=.46\textwidth]{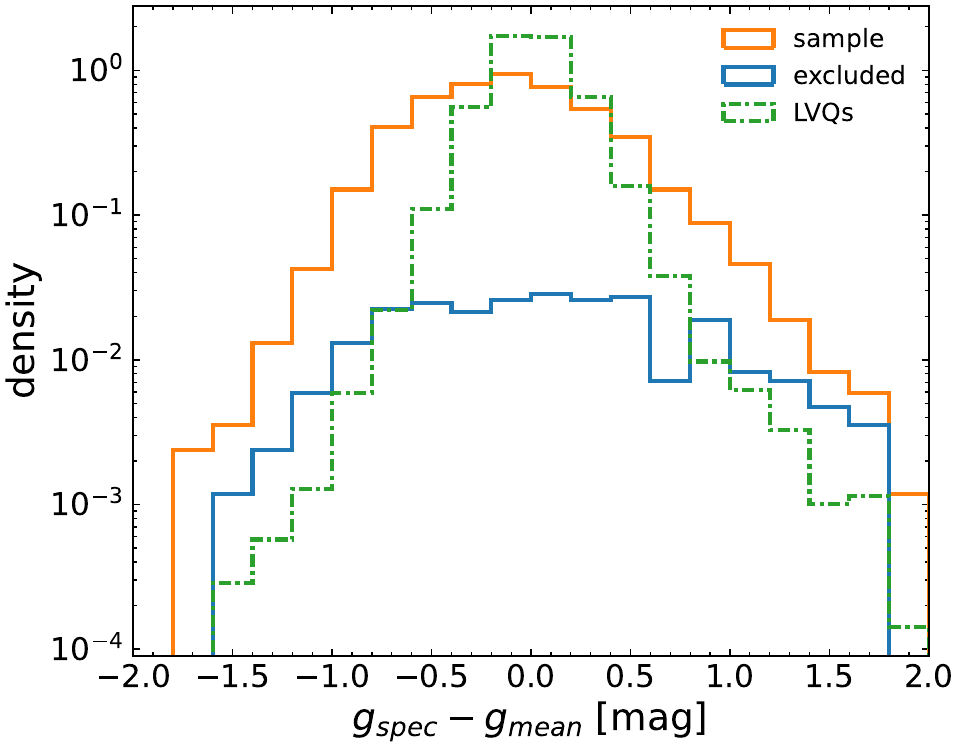}
   \caption{Normalized distribution of $g_{\text{spec}} - g_{\text{mean}}$ for multi-spectra EVQs (yellow) and LVQs (green). The blue solid line plots the excluded portion of the yellow line. A logarithmic y-axis is used to better visualize the outliers, i.e., with large $|g_{\text{spec}} - g_{\text{mean}}|$.
   \label{fig:A1_outlier}}
\end{figure}

In Fig. \ref{fig:A1_outlier}, we show the normalized distribution of $g_{\text{spec}} - g_{\text{mean}}$ (the difference between the $g$-band magnitude from the spectrum and the light curve mean, see \S3.1 in \citetalias{Ren2022} for details) for our sample and the LVQs. The distribution of $g_{\text{spec}} - g_{\text{mean}}$ of LVQs (for which we only expect small deviation from zero) does span a broad range, indicating the calibration issues are not negligible at large $|g_{\text{spec}} - g_{\text{mean}}|$. For our EVQs, our screening process does exclude considerable fraction of epochs at large $|g_{\text{spec}} - g_{\text{mean}}|$ but minor fraction at small $|g_{\text{spec}} - g_{\text{mean}}|$ (the ratio of blue to yellow line in Fig. \ref{fig:A1_outlier} could represent the excluded fraction as a function of $g_{\text{spec}} - g_{\text{mean}}$). At large $|g_{\text{spec}} - g_{\text{mean}}|$ (e.g. $>$ 0.8), we have excluded relatively more epochs than the prediction from LVQs (blue line lying above the green line), suggesting our approach is somehow conservative that some good spectra could have been excluded (which is however unavoidable).

\bibliography{sample631}{}
\bibliographystyle{aasjournal}

\end{document}